\documentclass[pra,twocolumn,floatfix,superscriptaddress,notitlepage,longbibliography]{revtex4-2}

\usepackage{changes}
\setdeletedmarkup{\color{blue}\sout{#1}}
\usepackage{cancel}

\usepackage[dvipsnames]{xcolor}
\usepackage[most]{tcolorbox}

\usepackage[normalem]{ulem}
\usepackage{graphicx, color, graphpap}
\usepackage{mathrsfs} 
\usepackage{enumitem}
\usepackage{amssymb}
\usepackage{amsthm}
\usepackage{multirow}
\usepackage[colorlinks=true,citecolor=blue,linkcolor=blue,urlcolor=blue]{hyperref}
\usepackage[T1]{fontenc}
\usepackage{verbatim}
\usepackage{mathtools}
\usepackage{titlesec}

\long\def\ca#1\cb{} 

\newcommand{\abs}[2][]{#1| #2 #1|}

\newcommand{\bramatket}[3]{\langle #1 \hspace{1pt} | #2 | \hspace{1pt} #3 \rangle}

\newcommand{\ket}[1]{|#1\rangle}               
\newcommand{\bra}[1]{\langle #1|}              
\newcommand{\dya}[1]{\ket{#1}\!\bra{#1}}

\newcommand{\Tr}{{\rm Tr}}

\newcommand{\ave}[1]{\langle #1\rangle} 
\renewcommand{\geq}{\geqslant}
\renewcommand{\leq}{\leqslant}

\newcommand{\ad}{^\dagger}

\newcommand*{\id}{\openone}

{}
{}
\usepackage{orcidlink}

\begin{document}
\title{Goldbach Representations in Quantum Optics and Metrology}
	
\author{Akira Sone
\orcidlink{0000-0003-3539-6140}}
\email{akira.sone@umb.edu}
\affiliation{Department of Physics, University of Massachusetts, Boston, Massachusetts 02125, USA}
\affiliation{The NSF AI Institute for Artificial Intelligence and Fundamental Interactions}

\author{Maxim Olshanii}
\affiliation{Department of Physics, University of Massachusetts, Boston, Massachusetts 02125, USA}

\begin{abstract}
We derive a relation between Goldbach representations and photon-number statistics in a quantum-optical setup. Considering an even-number Fock state injected into a stochastic beam splitter with uniformly distributed transmissivity and projections of the two output modes onto prime subspaces,  we demonstrate that each successful outcome corresponds to an Goldbach representation and that the averaged detection probability is exactly proportional to the number of Goldbach representations. Then, employing the Hardy--Littlewood asymptotic formula, we demonstrate that the post-measurement state generated by a beam splitter with fixed transmissivity exhibits standard quantum limit scaling of the ultimate precision limit in terms of phase estimation, whereas an equally weighted  superposition of Goldbach representations exhibits Heisenberg scaling. These results provide a quantum-optical realization of Goldbach representations and relate their statistics of with quantum metrology.  
\end{abstract}

\maketitle
\section{Introduction}
\label{sec:intro}
The story of Goldbach's conjecture began nearly three centuries ago, with an exchange of letters between Christian Goldbach to Leonhard Euler on June 7 and June 30, 1742. From this correspondence emerged what is now known as Goldbach's conjecture~\cite{crandall2005prime,pollack2009not}: every even integer greater than two can be expressed as the sum of two primes, namely,
\begin{align}
    \forall n\in\mathbb{N}_{\geq 2},~
    \exists\,p,q\in\mathbb{P}
    \quad\text{s.t.}\quad
    2n=p+q,
\end{align}
where $\mathbb{N}_{\geq 2}$ denotes the set of integers greater than or equal to $2$, and $\mathbb{P}$ denotes the set of all primes.

Despite its remarkably simple formulation, the conjecture has remained unsolved for more than 280 years, making it one of the oldest and most celebrated open problems in number theory, which has challenged generations of mathematicians and inspired several major breakthroughs in analytic number theory. In 1923, Hardy and Littlewood proposed an asymptotic formula for the number of Goldbach representations via the so-called circle method~\cite{hardy1923some}, providing a quantitative prediction for how frequently such prime decompositions should occur. In 1937, Vinogradov achieved a major breakthrough by proving that every sufficiently large odd integer can be expressed as the sum of three primes~\cite{vinogradov1937representation}. Around the same period, employing Vinogradov's methods, Chudakov, van der Corput, and Estermann independently proved that almost all even integers can be expressed as the sum of two primes~\cite{chudakov1937goldbach,vandercorput1936goldbach,estermann1938goldbach}. 

Progress toward the original binary conjecture continued with Chen's celebrated theorem in 1973, showing that every sufficiently large even integer can be written as the sum of a prime and a number having at most two prime factors via sieve methods~\cite{chen1973goldbach}. More recently, Helfgott completed the proof of the Goldbach's weak conjecture, establishing that every odd number greater than five is the sum of three primes~\cite{helfgott2013ternary,helfgott2015ternary}. Meanwhile, extensive numerical computations have verified the binary Goldbach conjecture for all even integers up to $4\times10^{18}$~\cite{oliveira2014goldbach}. However, after nearly three centuries of mathematical effort, and despite overwhelming analytical and numerical evidence, the original binary Goldbach conjecture remains unproven.

Goldbach's conjecture has also inspired research in quantum physics, where various approaches have been proposed to explore the properties of prime numbers by mapping number-theoretic problems onto quantum states and dynamics. For example, Sanchis-Lozano \textit{et al.} studied Goldbach's conjecture from the perspective of quantum field theory by introducing fermionic fields associated with prime numbers~\cite{sanchis2012prime}. Latorre and Sierra introduced quantum states that encode prime numbers and explored their applications to Goldbach's conjecture~\cite{latorre2014quantum}. They also investigated the entanglement properties of these states and their connection to prime distributions~\cite{latorre2015entanglement}, followed by further studies by García-Martín et al.~\cite{garciamartin2020prime}. More recently, Marchukov \textit{et al.} proposed a quantum search method based on driven atoms to identify Goldbach partitions~\cite{marchukov2025goldbach}, while Cassettari \textit{et al.} studied number-theoretic structures using atomic systems with specially designed potentials~\cite{cassettari2026quantum}. These studies show how quantum systems can offer new ways to explore problems in number theory. Most of them, however, rely on quantum states that encode prime numbers, quantum algorithms, or specially designed atomic systems.

In this paper, we take a different approach by considering a stochastic simple beam splitter (BS), whose photon-number statistics naturally describe the partitions of an integer. By dealing with the BS transmissivity (or reflectivity) as a uniformly distributed random variable, we establish a direct connection between the numbers of Goldbach representations and measurable photon-counting statistics. We then employ the Hardy–Littlewood asymptotic formula to study the contribution of the numbers of Goldbach representations to the quantum metrology~\cite{giovannetti2006quantum,giovannetti2006quantum,pirandola2018advances,degen2017quantum} by focusing on the scaling of quantum Fisher information (QFI)~\cite{liu2015quantum, liu2020quantum, Hayashi_2004,koczor2020variational, beckey2020variational, cerezo2021sub, sone2020generalized} for phase estimation, which quantifies the ultimate estimation precision limit characterized by the quantum Cram\'{e}r-Rao bound. This elucidates a relation between the statistical properties of prime numbers and the ultimate precision limits of quantum parameter estimation.

This paper is organized as follows. In Sec.~\ref{sec:setup}, we introduce the framework for mapping Goldbach's conjecture onto a quantum-optical system consisting of a beam splitter (BS) and projective measurements onto prime states. Then, in Sec.~\ref{sec:result}, we present our main results by establishing the connection between Goldbach representations and photon-counting statistics for even-number input Fock states. We further introduce the post-measurement state, which we call the beam-splitter (BS) Goldbach state, and investigate its metrological properties through phase estimation. In particular, we compare the scaling of the QFI for the BS Goldbach state with that of the equally weighted Goldbach state, followed by the concluding remarks in Sec.~\ref{sec:conc}. Throughout this paper, we set $\hbar=1$ and write $\id$ as the identity operator. 

\section{Setup}
\label{sec:setup}
\begin{figure}[htp!]
\centering
\includegraphics[width=1\linewidth]{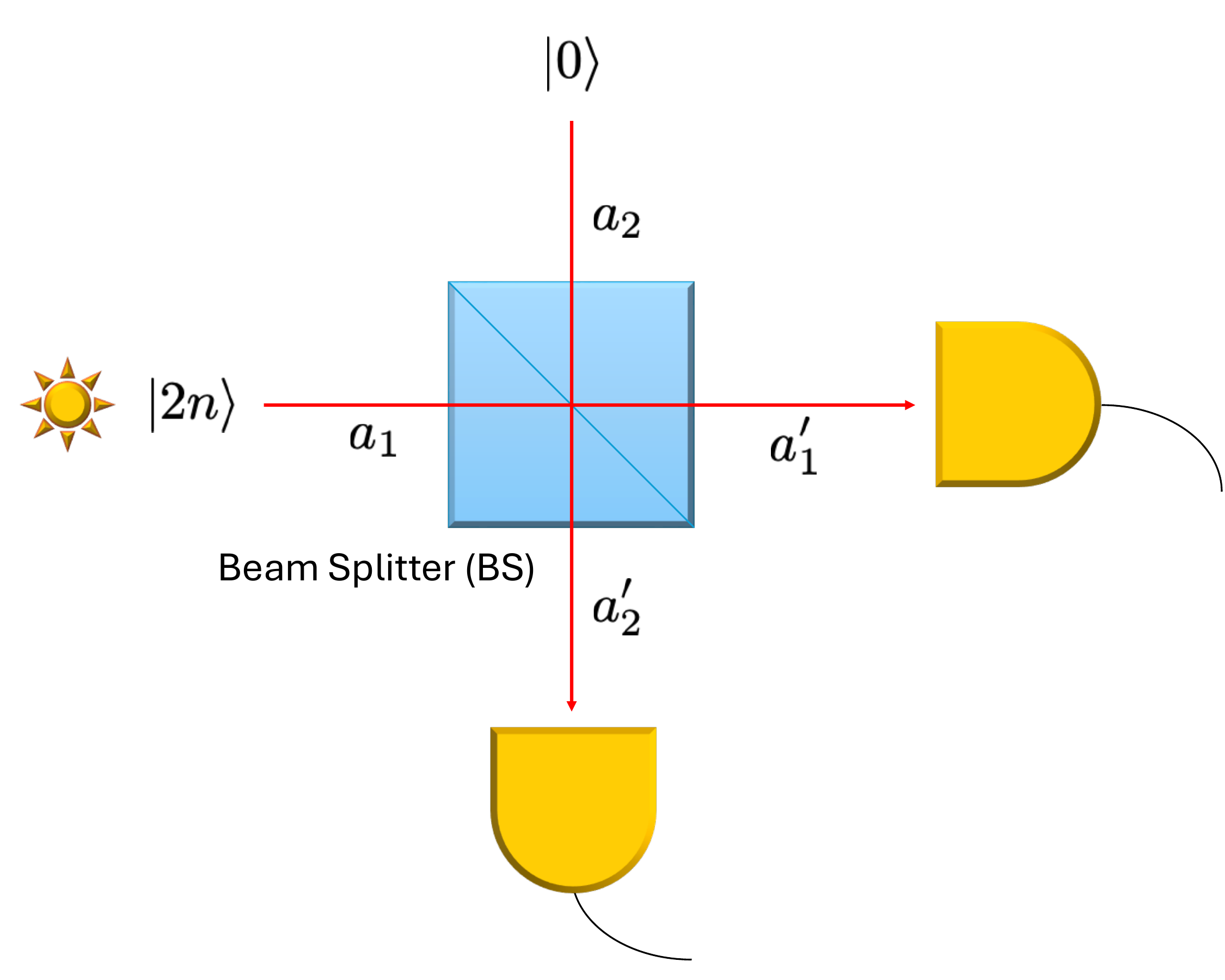}
    \caption{Splitting of $2n$ photons by a stochastic beam splitter (BS) with a random transmissivity $0\leq T \leq1$ uniformly distributed. Here, $a_1$ ($a_1^{\prime\dagger}$) and $a_2$ ($a_2^{\prime\dagger}$) denote the input (output) field operators for modes 1 and 2, respectively. An even-number Fock state $\ket{2n}$ is injected into mode 1, while mode 2 is initialized in the vacuum state $\ket{0}$, resulting in the output state $\ket{\psi_{2n,T}}$. }
\label{fig:BS}
\end{figure}

Let us consider a stochastic beam splitter (BS) with a random transmissivity 
\begin{align}
0\leq T \leq1
\end{align}
distributed uniformly. Denoting the input field operators of the two modes by $a_1$ and $a_2$, and the corresponding output field operators by $a_1'$ and $a_2'$, the unitary transformation $U_T$ describing the BS is given by~\cite{campos1989quantum,prasad1987quantum,fearn1987quantum,gerry2005introductory}
\begin{align}
\begin{split}
    \begin{pmatrix}
        a_1'\\
        a_2'
    \end{pmatrix}
    &=
    U_T\ad
    \begin{pmatrix}
        a_1\\
        a_2
    \end{pmatrix}
    U_T\\
    &=
    \begin{pmatrix}
        \sqrt{T}&-e^{-i\phi}\sqrt{1-T}\\
        e^{-i\phi}\sqrt{1-T}&\sqrt{T}
    \end{pmatrix}
    \begin{pmatrix}
        a_1\\
        a_2
    \end{pmatrix},
\end{split}
\label{eq:BS}
\end{align}
where $\phi\in\mathbb{R}$ denotes an arbitrary phase. 
We now inject an even-number Fock state into mode 1 and the vacuum state into mode 2,
\begin{align}
\ket{2n}_1\otimes\ket{0}_2\equiv\ket{2n,0}~~(n=2,3,4,\cdots).
\end{align}
The resulting output state is then given by (see Fig.~\ref{fig:BS})
\begin{align}
\ket{\psi_{2n,T}}
=
U_T\ket{2n,0}
=
\sum_{k=0}^{2n}A_{2n,T}(k)\ket{k,2n-k},
\end{align}
where the probability amplitude $A_{2n,T}(k)$ is defined as 
\begin{align}
A_{2n,T}(k)
\equiv e^{-i(2n-k)\phi}
\sqrt{\begin{pmatrix}
2n\\
k
\end{pmatrix}}
T^{k/2}
(1-T)^{n-k/2}.
\end{align}

\section{Main Results}
\label{sec:result}
Here, we present our main results by considering the projector onto the prime subspace, which we call prime projectors, as~\footnote{Note that for the input state $\ket{2n,0}$, the prime projector $\Pi_{\mathbb{P}}$ can be truncated at the largest prime less than $2n$. Nevertheless, we retain the general definition of $\Pi_{\mathbb{P}}$ throughout this paper to accommodate input states with indefinite photon numbers, such as the single-mode squeezed vacuum considered in Appendix~\ref{app:squeezed}.}
\begin{align}
    \Pi_{\mathbb{P}}\equiv\sum_{p\in\mathbb{P}}\dya{p}.
\end{align}
We define $\mathbb{G}_{2n}$ as the set of primes satisfying the Goldbach representations of $2n$,
\begin{align}
    \mathbb{G}_{2n}
    \equiv
    \left\{
    p\in\mathbb{P}\,\middle|\,2n-p\in\mathbb{P}
    \right\}\,~(n=2,3,4,\cdots).
\end{align}
Here, note that our definition of $\mathbb{G}_{2n}$ already accounts for the distinction between the ordered representations $p+q$ and $q+p$, which is different from Goldbach partitions. 
If we define $\abs{\mathbb{G}_{2n}}$ as the number of such primes $p$, then it is obvious that $\abs{\mathbb{G}_{2n}}$ is also equivalent to the number of Goldbach representations~\footnote{For example, for $2n=10$, we have $10=5+5=3+7=7+3$. Thus, the number of primes satisfying Goldbach's conjecture and equivalently the number of Goldbach representations for $2n=10$ are $\abs{\mathbb{G}_{2n}}=3$, where we distinguish the ordered representations $3+7$ and $7+3$. }.

\subsection{Goldbach Representations via Photon Counting}

Applying the tensor-product projector onto the prime subspaces of the two output modes yields
\begin{align}
    (\Pi_{\mathbb{P}}\otimes\Pi_{\mathbb{P}})\ket{\psi_{2n,T}}
    =
    \sum_{p\in\mathbb{G}_{2n}}
    A_{2n,T}(p)\ket{p,2n-p},
\end{align}
where each state $\ket{p,2n-p}$ encodes a Goldbach representation of $2n$. If Goldbach's conjecture were to fail for a given even integer $2n$, then $\mathbb{G}_{2n}=\emptyset$, and consequently $(\Pi_{\mathbb{P}}\otimes\Pi_{\mathbb{P}})\ket{\psi_{2n,T}}=0$.
The total probability of obtaining a Goldbach representation conditioned on $T$ is therefore
\begin{align}
\begin{split}
P_{2n}(T)
&\equiv
\bramatket{\psi_{2n,T}}
{(\Pi_{\mathbb{P}}\otimes\Pi_{\mathbb{P}})}
{\psi_{2n,T}}\\
&=\sum_{p\in\mathbb{G}_{2n}}\abs{A_{2n,T}(p)}^2,
\end{split}
\end{align}
where 
\begin{align}
\abs{A_{2n,T}(p)}^2=
\begin{pmatrix}
    2n\\
    p
\end{pmatrix}
T^{p}(1-T)^{2n-p}.
\end{align}
The phase $\phi$ does not appear in this probability because
\begin{align}
\abs{e^{-i(2n-p)\phi}}=1.
\end{align}
Here, because $T$ is uniformly distributed and 
\begin{align}
\int_{0}^{1}T^{p}(1-T)^{2n-p}dT = \frac{1}{2n+1}\frac{1}{\begin{pmatrix}2n\\p\end{pmatrix}},
\end{align}
by the law of total probability, the probability $P_{2n}$ that the even natural number $2n$ is represented by the sum of two ordered prime numbers, which we call Goldbach probability, is given by 
\begin{align}
P_{2n}\equiv\int_{0}^{1}P_{2n}(T)dT=\frac{\abs{\mathbb{G}_{2n}}}{2n+1},
\label{eq:P_{2n}}
\end{align}
where $2n+1$ is the total number of representations in the sum of two non-negative integers~\footnote{For example, $10$ has $2\times 5+1=11$ ways to represent, i.e., $10=0+10,~1+9,~2+8,~ 3+7,~4+6,~5+5,~6+4,~7+3,~8+2,~9+1,~10+0$. }.

Note that by defining $\rho_{2n}$ as 
\begin{align}
    \rho_{2n}\equiv \int_{0}^{1} U_{T}\dya{2n,0}U_{T}\ad dT,
\end{align}
$\rho_{2n}$ becomes a density matrix describing a quantum state; therefore, $P_{2n}$ is a measurable probability,
$P_{2n}$ is a measurable probability
\begin{align}
P_{2n}=\Tr\left[(\Pi_{\mathbb{P}}\otimes\Pi_{\mathbb{P}})\rho_{2n}\right],
\end{align}
so that the number of Goldbach representations can also be measurable, 
\begin{align}
    \abs{\mathbb{G}_{2n}}=(2n+1)\Tr\left[(\Pi_{\mathbb{P}}\otimes\Pi_{\mathbb{P}})\rho_{2n}\right].
\end{align}
This allows one to map the quantitative Goldbach conjecture on a measurable platform with photon counting.

\subsection{Metrological Scaling from Goldbach Prime Statistics}

\subsubsection{Case 1: Standard Quantum Limit}
First, assuming that Goldbach's conjecture holds and fixing the BS transmissivity $T$, we consider the post-measurement state, which we call beam-splitter (BS) Goldbach state,
\begin{align}
\begin{split}
\ket{\Phi_{2n,T}}
&\equiv
\frac{\Pi_{\mathbb{P}}\otimes\Pi_{\mathbb{P}}}
{\sqrt{P_{2n}(T)}}
\ket{\psi_{2n,T}}\\
&=
\sum_{p\in\mathbb{G}_{2n}}
\lambda_{2n,T}(p)\ket{p,2n-p},
\end{split}
\label{eq:goldbach_state}
\end{align}
where
\begin{align}
\lambda_{2n,T}(p)
\equiv
\frac{A_{2n,T}(p)}{\sqrt{P_{2n}(T)}},
\end{align}
with the normalization condition
\begin{align}
\sum_{p\in\mathbb{G}_{2n}}
\abs{\lambda_{2n,T}(p)}^2
=1.
\end{align}
We then apply the phase shift
\begin{align}
S(\varphi)=e^{i\varphi H},
\label{eq:phase_shift}
\end{align}
generated by
\begin{align}
H\equiv a_{1}\ad a_{1}-a_{2}\ad a_{2}.
\label{eq:hamiltonian}
\end{align}
The QFI for estimating $\varphi$ from $S(\varphi)\ket{\Phi_{2n,T}}$ is
\begin{align}
I_{\Phi}^{(\varphi)}(2n)
=
4\text{Var}_{\Phi}(H),
\end{align}
where
\begin{align}
\!\!\!\!\text{Var}_{\Phi}(H)
\equiv
\bramatket{\Phi_{2n,T}}{H^2}{\Phi_{2n,T}}
-
\bramatket{\Phi_{2n,T}}{H}{\Phi_{2n,T}}^2.
\end{align}
Using Eqs.~\eqref{eq:goldbach_state} and \eqref{eq:hamiltonian}, we obtain
\begin{align}
I_{\Phi}^{(\varphi)}(2n)
=16\text{Var}_{\lambda}(p)=
16\left(
\ave{p^2}_{\lambda}
-
\ave{p}_{\lambda}^{2}
\right),
\label{eq:QFI_Goldbach}
\end{align}
where, for any function $f(p)$, we define
\begin{align}
\ave{f(p)}_{\lambda}
\equiv
\sum_{p\in\mathbb{G}_{2n}}
\abs{\lambda_{2n,T}(p)}^2 f(p).
\end{align}

Next, we analyze the asymptotic scaling of $I_{\Phi}^{(\varphi)}(2n)$ for large $2n$. The probability distribution
$\abs{\lambda_{2n,T}(p)}^2$ is explicitly given by
\begin{align}
\begin{split}
\abs{\lambda_{2n,T}(p)}^2
&=
\frac{\abs{A_{2n,T}(p)}^2}
{\displaystyle
\sum_{p\in\mathbb{G}_{2n}}
\abs{A_{2n,T}(p)}^2}\\
&=
\frac{
\begin{pmatrix}
2n\\p
\end{pmatrix}
T^{p}(1-T)^{2n-p}
}{\displaystyle{\sum_{p\in\mathbb{G}_{2n}}
\begin{pmatrix}
2n\\p
\end{pmatrix}
T^{p}(1-T)^{2n-p}}}
\end{split}
\end{align}
By the central limit theorem, the binomial distribution approaches a Gaussian for asymptotically large $2n$,
\begin{align}
\begin{pmatrix}
2n\\
p
\end{pmatrix}
T^p(1-T)^{2n-p}
\sim
\frac{1}{\sqrt{2\pi\sigma^2}}
\exp\left[
-\frac{(p-\mu)^2}{2\sigma^2}
\right],
\end{align}
with
\begin{align}
\mu=2Tn,~\sigma=\sqrt{2T(1-T)n}.
\end{align}
Thus, $\abs{A_{2n,T}(p)}^2$ is sharply concentrated around
\begin{align}
p
=
2Tn+O(\sqrt{n}).
\end{align}

To evaluate the denominator, we employ the Hardy--Littlewood asymptotic formula~\cite{hardy1923some,crandall2005prime,pollack2009not},
\begin{align}
\abs{\mathbb{G}_{2n}}
\sim
C\int_{2}^{2n-2}
\frac{du}{\ln(u)\ln(2n-u)}
\sim
C\frac{2n}{[\ln(2n)]^2},
\end{align}
where we defined
\begin{align}
\!\!\!\! C
\equiv
2
\prod_{\substack{w>2\\w\in\mathbb{P}}}
\left(
1-\frac{1}{(w-1)^2}
\right)
\prod_{\substack{p>2,~p\in\mathbb{P}\\p\mid 2n}}
\left(\frac{p-1}{p-2}
\right),
\end{align}
where
\begin{align}
\prod_{\substack{w>2\\w\in\mathbb{P}}}
\left(
1-\frac{1}{(w-1)^2}
\right)=  0.6601618158\cdots   
\end{align}
is the so-called twin prime constant. 
More specifically, we assume the refined Hardy--Littlewood density for weighted Goldbach sums,
\begin{align}
dR
\sim
\frac{C}{\ln(u)\ln(2n-u)}\,du.
\end{align}
Under this assumption, the sum of a sufficiently smooth function $f(u)$ over the Goldbach representations can be approximated asymptotically by
\begin{align}
\sum_{p\in\mathbb{G}_{2n}}f(p)
\sim
C\int_{2}^{2n-2}
\frac{f(u)}{\ln(u)\ln(2n-u)}\,du.
\end{align}
It follows that the denominator can be approximately written as 
\begin{align}
\sum_{p\in\mathbb{G}_{2n}}
\abs{A_{2n,T}(p)}^2
\sim
C\int_{2}^{2n-2}
\frac{\abs{A_{2n,T}(u)}^2}
{\ln(u)\ln(2n-u)}
\,du.
\end{align}

Since the Gaussian approximation to $\abs{A_{2n,T}(u)}^2$ is sharply localized around
\begin{align}
u
=
2Tn+O(\sqrt{n})
\end{align}
with a width of order $O(\sqrt{n})$, we have
\begin{align}
\ln(u)
&=
\ln(2Tn)
+
O\left(\frac{1}{\sqrt{n}}\right),
\\
\ln(2n-u)
&=
\ln\bigl(2(1-T)n\bigr)
+
O\left(\frac{1}{\sqrt{n}}\right).
\end{align}
Therefore, to leading order, the denominator can be approximately written as 
\begin{align}
\begin{split}
\sum_{p\in\mathbb{G}_{2n}}&
\abs{A_{2n,T}(p)}^2\\
&\sim
\frac{C}
{\ln(2Tn)\ln\bigl(2(1-T)n\bigr)}
\int_{2}^{2n-2}
\abs{A_{2n,T}(u)}^2\,du
\\
&\sim
\frac{C}
{\ln(2Tn)\ln\bigl(2(1-T)n\bigr)}
\int_{-\infty}^{\infty}
\abs{A_{2n,T}(u)}^2\,du
\\
&\sim
\frac{C}
{\ln(2Tn)\ln\bigl(2(1-T)n\bigr)},
\end{split}
\end{align}
where we have used
\begin{align}
\int_{2}^{2n-2}
\abs{A_{2n,T}(u)}^2\,du
\sim
\int_{-\infty}^{\infty}
\abs{A_{2n,T}(u)}^2\,du
\sim 1.
\end{align}
This approximation follows because, for fixed $0\leq T\leq 1$ and asymptotically large $2n$, the distribution
$\abs{A_{2n,T}(u)}^2$ is sharply localized around $u = 2Tn$ inside the interval $u\in[2,2n-2]$.
Consequently, this does not modify the leading-order Gaussian envelope, and we may write
\begin{align}
\abs{\lambda_{2n,T}(p)}^2
\sim
\xi(p)
\propto
\exp\left[
-\frac{(p-\mu)^2}{2\sigma^2}
\right].
\end{align}
Therefore, to leading order, the variance of $p$ remains
\begin{align}
\text{Var}_{\lambda}(p)
\sim
2T(1-T)n.
\end{align}
Hence, from Eq.~\eqref{eq:QFI_Goldbach}, for fixed $T$, the asymptotic scaling of the QFI is
\begin{align}
I_{\Phi}^{(\varphi)}(2n)\sim 32T(1-T)n.
\end{align}
Then, the quantum Cram\'{e}r--Rao bound on the uncertainty of an unbiased estimator of $\varphi$ is given by
\begin{align}
\Delta\varphi \geq \frac{1}{\sqrt{I_{\Phi}^{(\varphi)}(2n)}}\sim \frac{1}{\sqrt{n}},.
\end{align}

corresponding to the standard quantum limit (SQL).

\subsubsection{Case 2: Heisenberg Limit}

Let us consider the equally weighted Goldbach state,
\begin{align}
    \ket{\Psi_{2n}}
    =\frac{1}{\sqrt{\abs{\mathbb{G}_{2n}}}}
    \sum_{p\in\mathbb{G}_{2n}}
    \ket{p,2n-p}.
\end{align}
Applying the phase shift in Eq.~\eqref{eq:phase_shift}, the QFI for estimating $\varphi$ from $S(\varphi)\ket{\Psi_{2n}}$ is given by
\begin{align}
    I(\varphi)=4\text{Var}_{\Psi}(H),
\end{align}
where
\begin{align}
    \text{Var}_{\Psi}(H)
    \equiv
    \bramatket{\Psi_{2n}}{H^2}{\Psi_{2n}}
    -\bramatket{\Psi_{2n}}{H}{\Psi_{2n}}^2.
\end{align}
Using Eq.~\eqref{eq:hamiltonian}, we obtain
\begin{align}
    I_{\Psi}^{(\varphi)}(2n)
    =16\text{Var}(p)
    =16\left(\ave{p^2}-\ave{p}^2\right),
\label{eq:QFI_Equally}
\end{align}
where, for any function $f(p)$, we define
\begin{align}
    \ave{f(p)}
    \equiv
    \frac{1}{\abs{\mathbb{G}_{2n}}}
    \sum_{p\in\mathbb{G}_{2n}}f(p).
\end{align}

We now evaluate the asymptotic scaling of $I_{\Psi}^{(\varphi)}(2n)$ for large $2n$.
Since 
\begin{align}
p\in\mathbb{G}_{2n}\,\Longrightarrow\, 2n-p\in\mathbb{G}_{2n},
\label{eq:implication}
\end{align}
we have
\begin{align}
    \sum_{p\in\mathbb{G}_{2n}}p
    =
    \sum_{p\in\mathbb{G}_{2n}}(2n-p).
\end{align}
Therefore,
\begin{align}
    2\sum_{p\in\mathbb{G}_{2n}}p
    =
    \sum_{p\in\mathbb{G}_{2n}}
    \left(p+(2n-p)\right)
    =
    2n\abs{\mathbb{G}_{2n}},
\end{align}
which gives
\begin{align}
    \ave{p}
    =
    \frac{1}{\abs{\mathbb{G}_{2n}}}
    \sum_{p\in\mathbb{G}_{2n}}p
    =n.
\label{eq:average}
\end{align}

For asymptotically large $2n$, we further have
\begin{align}
\!\!\!\!\ave{p^2}=
\frac{1}{\abs{\mathbb{G}_{2n}}}
\sum_{p\in\mathbb{G}_{2n}}p^2
\sim
\frac{
    \displaystyle
    \int_{2}^{2n-2}
    \frac{u^2}{\ln(u)\ln(2n-u)}du
    }{
    \displaystyle
    \int_{2}^{2n-2}
    \frac{1}{\ln(u)\ln(2n-u)}du
    }.
\label{eq:p^2_scaling}
\end{align}
In the asymptotic limit $n\to\infty$, the leading order becomes (See Appendix~\ref{app:leading-order} for detailed derivation) 
\begin{align}
\int_{2}^{2n-2}
    \frac{u^k}{\ln(u)\ln(2n-u)}\,du
    \sim
    \frac{(2n)^{k+1}}{(k+1)(\ln(2n))^2}\,~(k\geq0),
\label{eq:leading-order}
\end{align}
which follows that
\begin{align}
    \ave{p^2}
    =
    \frac{1}{\abs{\mathbb{G}_{2n}}}
    \sum_{p\in\mathbb{G}_{2n}}p^2
    \sim
    \frac{4}{3}n^2.
\label{eq:average^2}
\end{align}
Therefore, from Eqs.~\eqref{eq:average} and \eqref{eq:average^2}, we obatin
\begin{align}
    \text{Var}(p)
    =
    \ave{p^2}-\ave{p}^2
    \sim
    \frac{n^2}{3}.
\end{align}
Finally, using Eq.~\eqref{eq:QFI_Equally}, the QFI of the equally weighted Goldbach state asymptotically scales as
\begin{align}
    I_{\Psi}^{(\varphi)}(2n)
    \sim
    \frac{16}{3}n^2.
\end{align}
Then, the quantum Cram\'{e}r--Rao bound on the uncertainty of an unbiased estimator of $\varphi$ is given by
\begin{align}
\Delta\varphi \geq \frac{1}{\sqrt{I_{\Psi}^{(\varphi)}(2n)}}\sim \frac{1}{n}
\end{align}
corresponding to the Heisenberg limit.

\subsection{Numerical Example}
\begin{figure}[htp!]
\centering
\includegraphics[width=.9\linewidth]{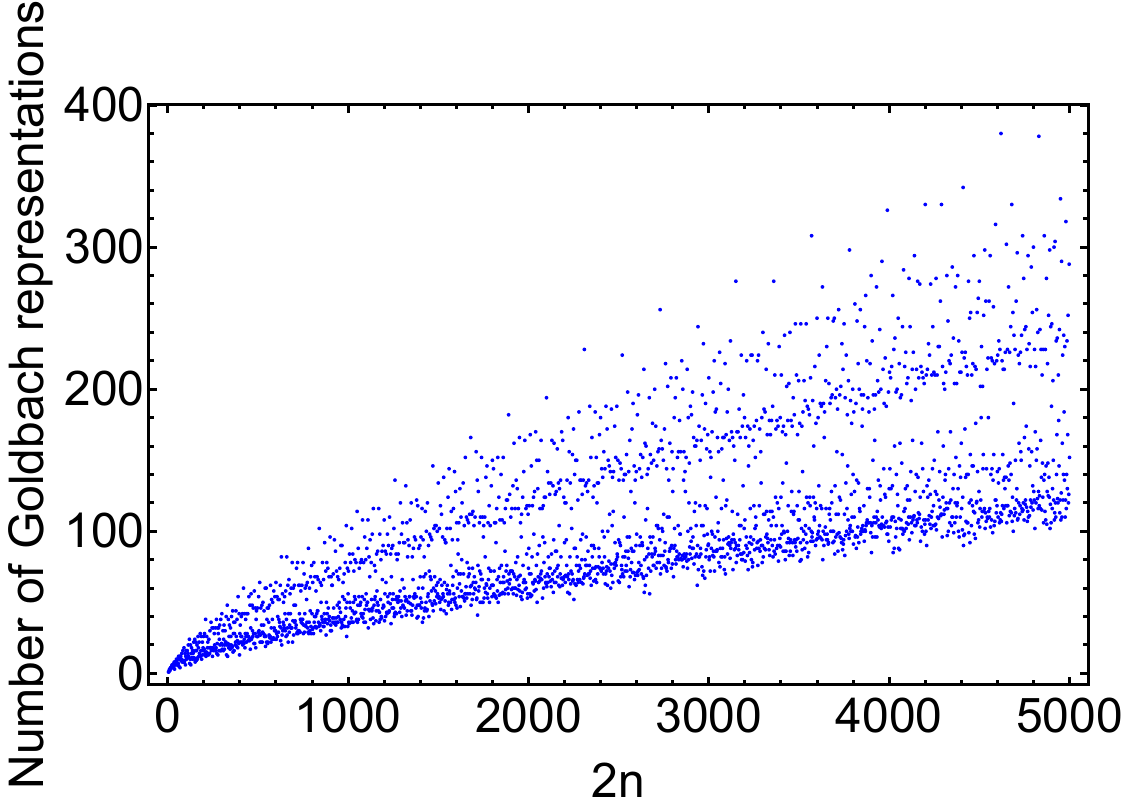}
    \caption{Plot of number of Goldbach representations $\abs{\mathbb{G}_{2n}}$ for $2n=4,6,8,\cdots,5000$.
}
\label{fig:Goldbach_G}
\end{figure}

\begin{figure}[htp!]
\centering
\includegraphics[width=1\linewidth]{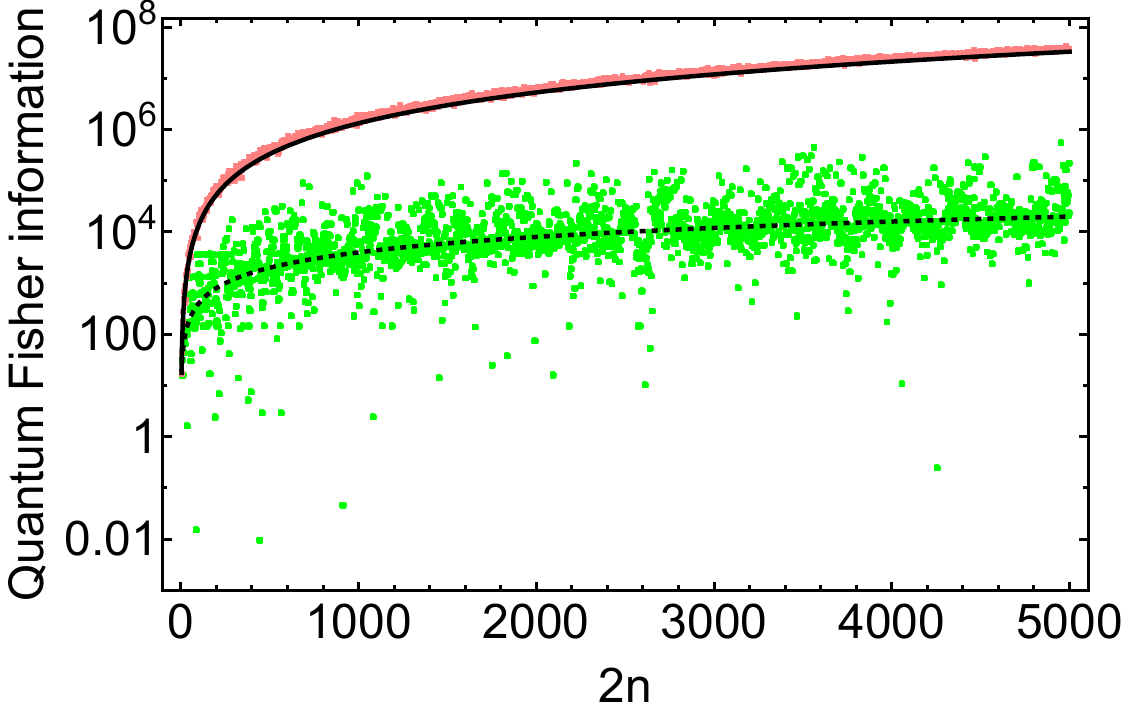}
    \caption{
Comparison of the QFI of the BS Goldbach state, $I_{\Phi}^{(\varphi)}(2n)$, at $T=1/2$, and the equally weighted Goldbach state, $I_{\Psi}^{(\varphi)}(2n)$, for $2n=4,6,8,\ldots,5000$ in the logarithmic scale. The green and pink dots represent $I_{\Phi}^{(\varphi)}(2n)$ and $I_{\Psi}^{(\varphi)}(2n)$, respectively. The black dashed line shows the asymptotic prediction $I_{\Phi}^{(\varphi)}(2n)\sim 8n$, while the black solid line represents $I_{\Psi}^{(\varphi)}(2n)\sim (16/3)n^2$. The equally weighted Goldbach state exhibits quadratic QFI scaling, in contrast to the linear scaling of the BS Goldbach state.
}
\label{fig:comparison}
\end{figure}

We consider the case of $50:50$ BS, i.e. $T=1/2$ for even-number Fock state of $2n=4,6,8,10,\cdots, 5000$. First, we illustrate the relation between $2n$ and $\abs{\mathbb{G}_{2n}}$ in Fig.~\ref{fig:Goldbach_G}. 
Then, we compare $I_{\Phi}^{(\varphi)}(2n)$ with $T=1/2$  and $I_{\Psi}^{(\varphi)}(2n)$ by plotting their relation with $2n$, which is illustrated in Fig.~\ref{fig:comparison}.
In Fig.~\ref{fig:comparison}, the green and pink dots represent $I_{\Phi}^{(\varphi)}(2n)$ and $I_{\Psi}^{(\varphi)}(2n)$, the QFIs of the BS Goldbach state and the equally weighted Goldbach state, respectively. 
The black dashed and solid line correspond to the linear asymptotic behavior $I_{\Phi}^{(\varphi)}(2n)\sim 8n$ and the quadratic asymptotic behavior $I_{\Psi}^{(\varphi)}(2n)\sim (16/3)n^2$ on a logarithmic scale, respectively. As shown in Fig.~\ref{fig:comparison}, the asymptotic predictions well align with the behaviors of the plots of the exact QFI values. These results indicate that a greater potential precision for phase estimation can be achieved in the Heisenberg limit with equally weighted Goldbach state.

For completeness, in Appendix~\ref{app:squeezed}, 
we also examine the case for the single-mode squeezed state, which is the superposition of the even-number Fock state.

\section{Conclusion}
\label{sec:conc}
We have established a quantum-optical framework that connects Goldbach representations with photon-number statistics. By injecting an even-number Fock state $\ket{2n}$ through a stochastic beam splitter and projecting the two output modes onto prime subspaces, we demonstrate that the output photon-number configurations correspond directly to Goldbach representations. Then, averaging the total probability of getting Goldbach representations over a uniformly distributed beam-splitter transmissivity, we show that the number of the Goldbach representations can be obtained from the measurable photon-counting probability. 

We further investigated the quantum metrological properties of quantum states constructed from Goldbach representations. For the post-measurement Goldbach state generated at fixed beam-splitter transmissivity, employing the Hardy--Littlewood asymptotic density, we demonstrate that the corresponding quantum Fisher information for phase estimation scales linearly with the photon number, exhibiting the scaling of the standard quantum limit. In contrast, with an equally weighted coherent superposition of all Goldbach representations, the corresponding quantum Fisher information grows quadratically with the photon number, leading to Heisenberg scaling.

These results show that the arithmetic structure of Goldbach representations can acquire a direct operational meaning in quantum optics. In particular, the metrological scaling is determined by how the different Goldbach representations are coherently weighted. Our framework therefore provides a simple connection between the statistical structure of Goldbach representations, photon-number statistics, and the ultimate precision limit, which can be studied within experimentally motivated quantum systems.

\acknowledgements{A.S. gratefully acknowledges his late grandfather, who introduced him to Goldbach's conjecture when he was in fifth grade and inspired his enduring interest in this longstanding mathematical problem. A.S also  acknowledges U.S. NSF under Grant No. OSI-2328774. A.S also acknowledges PHY-2425180 and Cooperative Agreement PHY-2019786. M.O.
was supported by the NSF Grant No. PHY-2309271.}

\appendix

\bibliography{ref.bib}

\onecolumngrid

\appendix

\section{Derivation of Eq.~\eqref{eq:leading-order}}
\label{app:leading-order}
We consider the integral
\begin{align}
\mathcal{I}_k(n)
\equiv
\int_2^{2n-2}
\frac{u^k}{\ln(u)\,\ln(2n-u)}\,du,~(k\geq 0).
\end{align}
Introducing the change of variable $u=2nx$, we obtain
\begin{align}
\mathcal{I}_k(n)
=\frac{(2n)^{k+1}}{\ln^2(2n)}
\int_{1/n}^{1-1/n}
\frac{x^k}
{\displaystyle{\left(1+\frac{\ln(x)}{\ln(2n)}\right)
 \left(1+\frac{\ln(1-x)}{\ln(2n)}\right)}}\,dx.
\end{align}
Let us define
\begin{align}
\mathcal{J}_k(n)
\equiv
\int_{1/n}^{1-1/n}
\frac{x^k}
{\displaystyle{\left(1+\frac{\ln(x)}{\ln(2n)}\right)
 \left(1+\frac{\ln(1-x)}{\ln(2n)}\right)}}\,dx.
\end{align}
Our objective is to show that
\begin{align}
\lim_{n\to\infty}\mathcal{J}_k(n)=\frac{1}{k+1}.
\label{eq:Jk-limit}
\end{align}

To establish this result, we divide the integration
interval into three regions:
\begin{align}
\begin{split}
L &:~\frac{1}{n}\leq x<\delta,\\
C &:~\delta\leq x\leq 1-\delta,\\
R &:~1-\delta<x\leq 1-\frac{1}{n},
\end{split}
\end{align}
where $0<\delta<1/2$ is fixed and $n$ is
sufficiently large  $n>1/\delta$.
We first take the asymptotic limit $n\to\infty$
with fixed $\delta$, followed by $\delta\to0$.

Accordingly, we decompose
\begin{align}
\mathcal{J}_k(n)
=\mathcal{J}_k^{(L)}(n,\delta)
+\mathcal{J}_k^{(C)}(n,\delta)
+\mathcal{J}_k^{(R)}(n,\delta),
\end{align}
where
\begin{align}
\mathcal{J}_k^{(L)}(n,\delta)
&\equiv
\int_{1/n}^{\delta}
\frac{x^k}
{\displaystyle{\left(1+\frac{\ln(x)}{\ln(2n)}\right)
 \left(1+\frac{\ln(1-x)}{\ln(2n)}\right)}}\,dx,\\
\mathcal{J}_k^{(C)}(n,\delta)
&\equiv
\int_{\delta}^{1-\delta}
\frac{x^k}
{\displaystyle{\left(1+\frac{\ln(x)}{\ln(2n)}\right)
 \left(1+\frac{\ln(1-x)}{\ln(2n)}\right)}}\,dx,\\
\mathcal{J}_k^{(R)}(n,\delta)
&\equiv
\int_{1-\delta}^{1-1/n}
\frac{x^k}
{\displaystyle{\left(1+\frac{\ln(x)}{\ln(2n)}\right)
 \left(1+\frac{\ln(1-x)}{\ln(2n)}\right)}}\,dx.
\end{align}

\subsection{Central region}
We first consider $C$, where
$\delta\leq x\leq1-\delta$.
For fixed $\delta$, both $\ln(x)$ and
$\ln(1-x)$ are bounded independently of $n$.
Consequently,
\begin{align}
\lim_{n\to\infty}\sup_{x\in[\delta,1-\delta]}
\left|\frac{\ln(x)}{\ln(2n)}\right|
&=0,\\
\lim_{n\to\infty}\sup_{x\in[\delta,1-\delta]}
\left|\frac{\ln(1-x)}{\ln(2n)}\right|
&=0
\end{align}
as $n\to\infty$.
Therefore, the integral converges uniformly
to $x^k$ on the region $C$, which yields
\begin{align}
\lim_{n\to\infty}\mathcal{J}_k^{(C)}(n,\delta)
=\int_\delta^{1-\delta}x^k\,dx
=\frac{(1-\delta)^{k+1}
-\delta^{k+1}}{k+1}.
\end{align}
Then, taking $\delta\to0$, we obtain
\begin{align}
\lim_{\delta\to0}
\lim_{n\to\infty}\mathcal{J}_k^{(C)}(n,\delta)
=\frac{1}{k+1}.
\label{eq:central-limit}
\end{align}

\subsection{Left Endpoint Region}
We next consider $L$, where
$1/n\leq x\leq\delta$.
Since $\ln(x)/\ln(2n)$ is not uniformly
small near $x=1/n$, the approximation
used in the central region cannot be
applied directly. Returning to the original variable
$u=2nx$, we have
\begin{align}
\mathcal{J}_k^{(L)}(n,\delta)
=\frac{\ln^2(2n)}{(2n)^{k+1}}
\int_2^{2n\delta}
\frac{u^k}{\ln(u)\,\ln(2n-u)}\,du.
\end{align}
For fixed $0<\delta<1/2$ and sufficiently
large $n$, the integration interval is
nonempty, and
\begin{align}
u\geq2\,,~
2n-u\geq2n(1-\delta)>1.
\end{align}
Hence, both logarithms are positive.
Moreover,
\begin{align}
\ln(2n-u)\geq\ln(2n(1-\delta)),
\end{align}
which implies
\begin{align}
0\leq \mathcal{J}_k^{(L)}(n,\delta)
\leq
\frac{\ln^2(2n)}
{(2n)^{k+1}\ln(2n(1-\delta))}
\int_2^{2n\delta}
\frac{u^k}{\ln(u)}\,du.
\label{eq:left-bound}
\end{align}

The remaining integral can be evaluated
asymptotically using L'H\^opital's rule:
 for $k\geq0$,
\begin{align}
\lim_{\alpha\to\infty}
\frac{\displaystyle\int_2^{\alpha}
\frac{u^k}{\ln(u)}\,du}
{\displaystyle\frac{\alpha^{k+1}}
{(k+1)\ln \alpha}}=
\lim_{\alpha\to\infty}
\frac{1}
{\displaystyle{1-\frac{1}{(k+1)\ln \alpha}}}
=1.
\end{align}
Therefore,
\begin{align}
\int_2^{2n\delta}
\frac{u^k}{\ln(u)}\,du
\sim
\frac{(2n\delta)^{k+1}}
{(k+1)\ln(2n\delta)}
\end{align}
as $n\to\infty$ with fixed $\delta$. Substituting this asymptotic expression
into Eq.~\eqref{eq:left-bound}, we find
\begin{align}
\begin{split}
\limsup_{n\to\infty}\mathcal{J}_k^{(L)}(n,\delta)
&\leq \frac{\delta^{k+1}}{k+1}
\lim_{n\to\infty}
\frac{\ln^2(2n)}
{\ln(2n(1-\delta))\ln(2n\delta)}\\
&=\frac{\delta^{k+1}}{k+1}\lim_{n\to\infty}\frac{1}{\displaystyle{\left(1+\frac{\ln(1-\delta)}{\ln(2n)}\right)\left(1+\frac{\ln(\delta)}{\ln(2n)}\right)}}
\\
&=\frac{\delta^{k+1}}{k+1}.
\end{split}
\end{align}
Then, taking
$\delta\to0$ gives
\begin{align}
\lim_{\delta\to0}
\limsup_{n\to\infty}\mathcal{J}_k^{(L)}(n,\delta)=0.
\label{eq:left-limit}
\end{align}
Because  $\mathcal{J}_k^{(L)}\geq0$, we can conclude
\begin{align}
    \lim_{\delta\to 0}\lim_{n\to\infty}\mathcal{J}_{k}^{(L)}(n,\delta)=0.
\end{align}

\subsection{Right Endpoint Region.}
Finally, we consider $R$, where
$1-\delta\leq x\leq1-1/n$.
Introducing $y=1-x$, we obtain
\begin{align}
\mathcal{J}_k^{(R)}(n,\delta)
=
\int_{1/n}^{\delta}
\frac{(1-y)^k\,dy}
{\displaystyle{\left(1+\frac{\ln(y)}{\ln(2n)}\right)
 \left(1+\frac{\ln(1-y)}{\ln(2n)}\right)}}.
\end{align}
For $1/n\leq y\leq\delta<1/2$
and sufficiently large $n>1/\delta$, both
denominator factors are positive:
\begin{align}
1+\frac{\ln(y)}{\ln(2n)}
&=\frac{\ln(2ny)}{\ln(2n)}>0,\\
1+\frac{\ln(1-y)}{\ln(2n)}
&=\frac{\ln(2n(1-y))}{\ln(2n)}>0.
\end{align}
Furthermore, since $k\geq0$,
\begin{align}
0<(1-y)^k\leq1.
\end{align}
It follows that
\begin{align}
0\leq \mathcal{J}_k^{(R)}(n,\delta)
\leq
\int_{1/n}^{\delta}
\frac{1}
{\displaystyle{\left(1+\frac{\ln(y)}{\ln(2n)}\right)
 \left(1+\frac{\ln(1-y)}{\ln(2n)}\right)}}\,dy=J_0^{(L)}(n,\delta).
\end{align}
Applying Eq.~\eqref{eq:left-limit}
with $k=0$, we conclude that
\begin{align}
\lim_{\delta\to0}
\limsup_{n\to\infty}\mathcal{J}_k^{(R)}(n,\delta)=0.
\label{eq:right-limit}
\end{align}
Because $\mathcal{J}_{k}^{(R)}(n,\delta)\geq0$, we can conclude
\begin{align}
\lim_{\delta\to0}
\lim_{n\to\infty}\mathcal{J}_k^{(R)}(n,\delta)=0.    
\end{align}

\subsection{Asymptotic Result}
Combining Eqs.~\eqref{eq:central-limit},
\eqref{eq:left-limit}, and
\eqref{eq:right-limit}, we conclude that
the endpoint contributions vanish in the
successive limits $n\to\infty$ and
$\delta\to0$, so that we can obtain 
\begin{align}
\lim_{n\to\infty}\mathcal{J}_k(n)=\frac{1}{k+1},
\end{align}
which results in 
\begin{align}
\mathcal{I}_k(n)\sim
\frac{(2n)^{k+1}}
{(k+1)\ln^2(2n)}\,,~
(k\geq 0).
\label{eq:Ik-asymptotic}
\end{align}

\section{Single-Mode Squeezed State Input}
\label{app:squeezed}

In this appendix, we extend our analysis of Goldbach-representation statistics and QFI to a single-mode squeezed vacuum state, which is a superposition of even-number Fock states~\cite{gerry2005introductory},
\begin{align}
    \ket{\xi_r}
    = \sum_{n=0}^{\infty}\alpha_{2n,r}\ket{2n},
\end{align}
where
\begin{align}
    \alpha_{2n,r}
    \equiv \frac{(-1)^n}{\sqrt{\cosh(r)}}
    \frac{\sqrt{(2n)!}}{2^n n!}
    e^{in\chi}\tanh^n(r)
\end{align}
with $\chi\in\mathbb{R}$. 
We consider the input state $\ket{\xi_r,0}$ incident on the beam splitter (BS) described by Eq.~\eqref{eq:BS}. 
The resulting two-mode output state is
\begin{align}
\begin{split}
    \ket{\psi_{r,T}}
    = U_T\ket{\xi_r,0}= \sum_{n=0}^{\infty}\alpha_{2n,r}
    \sum_{k=0}^{2n}A_{2n,T}(k)\ket{k,2n-k}.
\end{split}
\end{align}

\subsection{Goldbach Representation Statistics}
To consider the Goldbach representations in this scenario, we need to include the case of $n=0,1$ for convenience; therefore, we have 
\begin{align}
\mathbb{G}_0=\mathbb{G}_2=\emptyset\,\Longrightarrow\,\abs{\mathbb{G}_{0}}=\abs{\mathbb{G}_{2}}=0,
\label{eq:emptyset}
\end{align}
so that we need to impose
\begin{align}
    A_{0,T}(p)=    A_{2,T}(p) =0.
\label{eq:zero_probability}
\end{align}
Then, the probability of detecting a Goldbach representation in the output modes is therefore
\begin{align}
\begin{split}
    P_r(T)&=\bramatket{\psi_{r,T}}{\left(\Pi_{\mathbb{P}}\otimes\Pi_{\mathbb{P}}\right)}{\psi_{r,T}}\\
    &= \sum_{n=2}^{\infty}\abs{\alpha_{2n,r}}^2
    \sum_{p\in\mathbb{G}_{2n}}|A_{2n,T}(p)|^2\\
    &= \sum_{n=2}^{\infty}\abs{\alpha_{2n,r}}^2 P_{2n}(T),
\end{split}
\label{eq:P_r}
\end{align}
where we have used Eqs.~\eqref{eq:emptyset} and \eqref{eq:zero_probability}.
Here, $\abs{\alpha_{2n,r}}^2$ is the probability that the input squeezed vacuum contains $2n$ photons, while $P_{2n}(T)$ is the conditional probability of detecting a Goldbach representation at the BS output, given a total photon number of $2n$.
Finally, averaging over the BS transmissivity $T$ yields
\begin{align}
\int_0^1 P_r(T)dT
    = \sum_{n=2}^{\infty}
    \abs{\alpha_{2n,r}}^2
    \frac{\abs{\mathbb{G}_{2n}}}{2n+1},
\end{align}
where we have used Eq.~\eqref{eq:P_{2n}}.

\subsection{Quantum Fisher Information}

For a fixed BS transmissivity $T$, the normalized post-measurement state becomes
\begin{align}
\ket{\Psi_{r,T}}
    \equiv
    \frac{\Pi_{\mathbb{P}}\otimes\Pi_{\mathbb{P}}}
    {\sqrt{P_r(T)}}\ket{\psi_{r,T}}
    =\sum_{n=2}^{\infty}\sum_{p\in\mathbb{G}_{2n}}
    \alpha_{2n,r}\frac{A_{2n,T}(p)}
{\sqrt{P_r(T)}}\ket{p,2n-p}.
\end{align}
Note that this state does not require to assume that Goldbach's conjecture is true because $P_{r}(T)\neq 0$ as we take into account all even numbers. 
Introducing the normalized amplitudes
\begin{align}
    \Omega_{r,T}(2n,p)
    \equiv
    \alpha_{2n,r}\frac{A_{2n,T}(p)}
    {\sqrt{P_r(T)}},
\end{align}
we can express the post-measurement state as
\begin{align}
    \ket{\Psi_{r,T}}
    =\sum_{n=2}^{\infty}\sum_{p\in\mathbb{G}_{2n}}
    \Omega_{r,T}(2n,p)\ket{p,2n-p}.
\end{align}
By Eq.~\eqref{eq:P_r}, these amplitudes satisfy the normalization condition
\begin{align}
    \sum_{n=2}^{\infty}\sum_{p\in\mathbb{G}_{2n}}
    \abs{\Omega_{r,T}(2n,p)}^2=1.
\end{align}

Applying the phase-shift operator $S(\varphi)$ defined in Eq.~\eqref{eq:phase_shift}, the quantum Fisher information (QFI) for estimating $\varphi$ from $S(\varphi)\ket{\Psi_{r,T}}$ is
\begin{align}
    I_{\Psi}^{(\varphi)}(r,T)
    =4\mathrm{Var}_{\Psi_{r,T}}(H),
\end{align}
where
\begin{align}
\mathrm{Var}_{\Psi_{r,T}}(H)\equiv
\bramatket{\Psi_{r,T}}{H^2}{\Psi_{r,T}}
-
\bramatket{\Psi_{r,T}}{H}{\Psi_{r,T}}^2.
\end{align}
Using the Hamiltonian in Eq.~\eqref{eq:hamiltonian}, we obtain
\begin{align}
    I_{\Psi}^{(\varphi)}(r,T)
    =16\mathrm{Var}_{\Omega}(p-n),
\end{align}
where the variance of an arbitrary function $f(p,n)$ with respect to the probability distribution $\abs{\Omega_{r,T}(2n,p)}^2$ is defined as
\begin{align}
    \mathrm{Var}_{\Omega}(f)
    \equiv
    \sum_{n=2}^{\infty}\sum_{p\in\mathbb{G}_{2n}}
    \abs{\Omega_{r,T}(2n,p)}^2 f^2(p,n)
    -
    \left(
    \sum_{n=2}^{\infty}\sum_{p\in\mathbb{G}_{2n}}
    \abs{\Omega_{r,T}(2n,p)}^2 f(p,n)
    \right)^2.
\end{align}
Particularly, when $T=1/2$, we have 
\begin{align}
    \abs{A_{2n,1/2}(p)}^2=\begin{pmatrix}
        2n\\
        p
    \end{pmatrix}
    \left(\frac{1}{2}\right)^{2n},
\end{align}
so that 
\begin{align}
 \abs{A_{2n,1/2}(p)}^2 = \abs{A_{2n,1/2}(2n-p)}^2.   
\end{align}
Therefore, we have
\begin{align}
\sum_{n=2}^{\infty}\sum_{p\in\mathbb{G}_{2n}}\abs{\Omega_{r,1/2}(2n,p)}^2(p-n)
=\frac{1}{P_{r}(T)}\sum_{n=2}^{\infty}\abs{\alpha_{2n,r}}^2\left(\frac{1}{2}\right)^{2n}\sum_{p\in\mathbb{G}_{2n}}
\begin{pmatrix}
    2n\\
    p
\end{pmatrix}
(p-n).
\end{align}
From Eq.~\eqref{eq:implication}, because  
\begin{align}
\begin{split}
2\sum_{p\in\mathbb{G}_{2n}}\begin{pmatrix}
    2n\\
    p
\end{pmatrix}(p-n)=&\sum_{p\in\mathbb{G}_{2n}}
\begin{pmatrix}
    2n\\
    p
\end{pmatrix}(p-n)+\sum_{2n-p\in\mathbb{G}_{2n}}\begin{pmatrix}
    2n\\
    2n-p
\end{pmatrix}(2n-p-n)\\
=&\sum_{p\in\mathbb{G}_{2n}}
\begin{pmatrix}
    2n\\
    p
\end{pmatrix}(p-n)+\sum_{p\in\mathbb{G}_{2n}}\begin{pmatrix}
    2n\\
    p
\end{pmatrix}(2n-p-n)\\
=&\sum_{p\in\mathbb{G}_{2n}}
\begin{pmatrix}
    2n\\
    p
\end{pmatrix}\Big((p-n)+(2n-p-n)\Big)\\
=&0,
\end{split}
\end{align}
we have 
\begin{align}
 \sum_{n=2}^{\infty}\sum_{p\in\mathbb{G}_{2n}}\abs{\Omega_{r,1/2}(2n,p)}^2(p-n)=0,   
\end{align}
so that the QFI can be simply given by 
\begin{align}
I_{\Psi}^{(\varphi)}(r,1/2)=16\sum_{n=2}^{\infty}\sum_{p\in\mathbb{G}_{2n}}\abs{\Omega_{r,1/2}(2n,p)}^2(p-n)^2.
\end{align}

\end{document}